\documentclass[11pt,a4paper]{article}
\usepackage{amsmath,amssymb,mathrsfs}
\usepackage[dvipdfmx]{graphicx}

\allowdisplaybreaks

\newcommand{\ms}[1]{\mathscr{#1}}
\newcommand{\mc}[1]{\mathcal{#1}}

\begin{document}

\begin{titlepage}

\begin{center}

\begin{flushright}
\hfill UT-14-12
\\
\hfill IPMU14-0090
\end{flushright}

\vskip 1.5in

{\LARGE \bf 

  Complexified Starobinsky Inflation in Supergravity\\
   in the Light of
  Recent BICEP2 Result

}

\vskip .5in

{\large
Koichi Hamaguchi$^{a,b}$, 
Takeo Moroi$^{a,b}$,
Takahiro Terada$^{a}$, 
}

\vskip 0.25in

{\em
${}^a$Department of Physics, The University of Tokyo,
Tokyo 113-0033, Japan\\
${}^b$Kavli Institute for the Physics and Mathematics of the Universe (Kavli IPMU),\\
WPI, TODIAS, The University of Tokyo, Chiba 277-8568, Japan
}

\end{center}
\vskip .5in

\begin{abstract}
  
  Motivated by the recent observation of the $B$-mode signal in the
  cosmic microwave background by BICEP2, we study the Starobinsky-type
  inflation model in the framework of old-minimal supergravity, where
  the inflaton field in the original (non-supersymmetric) Starobinsky
  inflation model is promoted to a complex field.  We study how the inflaton
  evolves on the two-dimensional field space, varying the initial
  conditions.  We show that (i) one of the scalar fields has a very
  steep potential once the trajectory is off from that of the original
  Starobinsky inflation, and that (ii) the $B$-mode signal observed by
  BICEP2 is too large to be consistent with the prediction of the
  model irrespective of the initial conditions.  Thus, the BICEP2
  result strongly disfavors the complexified Starobinsky inflation in
  supergravity.

\end{abstract}

\end{titlepage}


Recently, $B$-mode polarization of cosmic microwave background (CMB)
has been observed by BICEP2, which indicated a large tensor-to-scalar
ratio of \cite{Ade:2014xna}
\begin{align}
  r^{\rm (BICEP2)} = 0.2^{+0.07}_{-0.05}.
  \label{r(BICEP2)}
\end{align}
The observation of BICEP2 provides a significant constraint on
inflationary models because the value of $r$ is directly related to
the scale of inflation (i.e., the expansion rate during inflation).
In particular, the BICEP2 result strongly disfavors one of the
interesting possibilities, i.e., Starobinsky inflation model
\cite{Starobinsky, Starobinsky2} which utilizes a scalar degree of
freedom in the gravitational sector as an inflaton.  This is because the
Starobinsky inflation predicts $r$ of the order of $10^{-3}$, which is
significantly smaller than the BICEP2 result.

If one extends the model, this conclusion may change.  
The extension we consider in the present study is to
supersymmetrize the model because supersymmetry is a prominent
candidate of the physics beyond the Standard Model.
In such a model, the inflation can still be realized solely by the
gravitational sector, while new scalar degrees of freedom are automatically introduced,
which may affect the dynamics of inflation.

The Starobinsky model is based on a modified theory of gravity, so we
need to consider a modified theory of supergravity.  There are two
minimal off-shell formulations of supergravity: the
old-minimal~\cite{Ferrara:1978em, Stelle:1978ye, Fradkin:1978jq} and
the new-minimal~\cite{Sohnius:1981tp} supergravity.  Supergravity
embedding of Starobinsky model has been studied both in the
old-minimal~\cite{Cecotti:1987sa, Ferrara:2013wka, Ketov:2013dfa} and the 
new-minimal~\cite{Cecotti:1987qe, Farakos:2013cqa} supergravities.
These studies share the original philosophy of the Starobinsky model
in the sense that the supergravity generalizations of the model rely
solely on (super)geometrical or (super)gravitational
quantities.\footnote{
In this respect, see Ref.~\cite{Ellis:2013zsa} for the inflationary
scenario induced by gravitino condensation.
 Closely related works to Refs.~\cite{Cecotti:1987sa, Ferrara:2013wka, Ketov:2013dfa, Cecotti:1987qe, Farakos:2013cqa}
  include Refs.~\cite{Kallosh:2013lkr, eno} in the old-minimal
  formulation and Refs.~\cite{fklp} (see also Refs.~\cite{fresorin}) in both formulations.
  See also other recent related works~\cite{SCDterm, Roest:2013aoa} in
  supergravity.  These can reproduce the scalar potential of the dual
  theory of the Starobinsky model~\cite{oldr}, but do not necessarily
  have pure (super)geometrical or (super)gravitational interpretation.
  Generalization of the duality~\cite{Cecotti:1987sa} between
  higher-curvature supergravity and standard matter-coupled
  supergravity has recently been discussed in
  Ref.~\cite{Cecotti:2014ipa} which provides the higher-curvature
  supergravity representation of the attractor
  model~\cite{Kallosh:2013yoa}.
}
~ The old-minimal realization of Starobinsky model is possible with
generic ``K\"{a}hler potential'' and ``superpotential'' of scalar
curvature supermultiplet with extra propagating scalar degrees of
freedom other than the inflaton (also called
scalaron)~\cite{Cecotti:1987sa, Ferrara:2013wka,
  Ketov:2013dfa}.\footnote
{ Imposing a constraint $\mc{R}^{2}=0$, one can construct the old-minimal
  higher-curvature supergravity with only one (pseudo)scalar in
  addition to the scalaron~\cite{Antoniadis:2014oya}.
  Even in this case, the discussion after Eq.~(\ref{eq:ReTImT}) holds.
    }
On the contrary, the new-minimal realization has a Higgsed (massive)
vector field as well as the inflaton~\cite{Cecotti:1987qe,
  Farakos:2013cqa}.  Thus, we consider the old-minimal
supergravity because it automatically introduces new scalar degrees of
freedom.

In this letter, we study the Starobinsky-type inflation model in the
framework of old-minimal supergravity.  We pay particular attention to
the fact that there exist two scalar degrees of freedom originating
from the gravity multiplet in such a model.  We study the evolution of
the inflaton on the two-dimensional field space.  We will see that the
potential of one of the scalar fields becomes very steep once the
trajectory is off from that of the original Starobinsky inflation.  We
also show that the tensor-to-scalar ratio in the supergravity
Starobinsky model is too small to be consistent with the BICEP2 result
even though the field space is enlarged.


The generic action of the old-minimal supergravity~\cite{Cecotti:1987sa, Ketov:2013dfa} is, in chiral curved superspace language,\footnote{Throughout this letter, we use the Planck unit $M_P=1$, where 
$M_P\simeq 2.4\times 10^{18}~\text{GeV}$ is the reduced Planck scale.
}
\begin{align} \label{cha}
S=\int \text{d}^{4}x \mathrm{d}^{2}\Theta 2 \ms{E} \left [ -\frac{1}{8}\left( \bar{\ms{D}}\bar{\ms{D}}-8{\mc R}\right) N({\mc R},\bar{\mc R})+
F({\mc R}) \right ] +\text{H.c.}
\end{align}
where $N(\mc{R},\bar{\mc{R}})$ and $F(\mc{R})$ are the hermitian and the holomorphic functions of the scalar curvature chiral superfield $\mc{R}$, respectively.
The superfield $\mc{R}$ contains Ricci scalar curvature $R$ in its $\Theta\Theta$ component and gravitino in its $\Theta$ and $\Theta\Theta$ components.
It also contains a complex scalar $M$ and real vector $b^{\mu}$.
These are auxiliary fields in the case of the minimal action with $N=-3$ and $F=0$.
For generic functions $N$ and $F$, however, these become dynamical.

The theory is classically equivalent~\cite{Cecotti:1987sa, Ketov:2013sfa} to the standard matter-coupled supergravity~\cite{Wess:1992cp}
\begin{align}
S=\int d^{4}xd^{2}\Theta 2\ms{E} \left [ \frac{3}{8}\left( \bar{\ms{D}}\bar{\ms{D}}-8\mc{R}\right)e^{-K/3}+W \right ] +\text{H.c.}
\end{align}
 with the following no-scale type K\"{a}hler potential and superpotential:
\begin{align}
K=&-3\ln\left( \frac{T+\bar{T}-N(S,\bar{S})}{3} \right) , \label{kahlerp} \\
W=&2TS+F(S) \label{superp}.
\end{align}
Linearized analysis of the original picture (higher-curvature supergravity) for a simple function $N(\mc{R},\bar{\mc{R}})$ has been performed in Ref.~\cite{Ferrara:1978rk}.
Bosonic Lagrangian of the original picture and comparisons of both pictures are described in Ref.~\cite{Ketov:2013dfa}.
Note that {\it any} $N$ and $F$ functions lead to the unique K\"{a}hler and superpotentials for $T$ because the origin of $T$ is a Lagrange multiplier.
In particular, canonically normalized field $X=\sqrt{3/2}\ln (1+2 \text{Re}T/3)$ 
along the real axis ($\text{Im}T=S=0$)  has the Starobinsky potential
(cf. Eq.~(\ref{eq:LXY})).
Roughly speaking, $\text{Re}T$, $\text{Im}T$, $S$, and $\bar{S}$ in this picture correspond to $R$, $\partial_{\mu}b^{\mu}$, $M$, and $\bar{M}$ in the original geometrical picture, respectively. 
In this letter, we focus on the standard matter-coupled supergravity picture.

Consider a K\"{a}hler potential for $S$,
\begin{align}
N(S,\bar{S}) =-3+\frac{12}{m^{2}} S\bar{S}-\zeta \left( S \bar{S} \right)^{2}.
\end{align}
The first term (constant) is needed to reproduce Einstein supergravity.
The second term leads to the kinetic term of the new degrees of freedom.
However, this term produces the scalar potential unbounded below in the region of large $|S|$.
Instability for radial $|S|$ direction is stabilized by the third term proportional to $\zeta$ (see {\textit e.g.} Refs.~\cite{Lee:2010hj,Ferrara:2010in,
Kallosh:2013lkr,Ketov:2013dfa} and references therein).  
Small $\zeta$ makes other local minima near the original minimum ($T=S=0$).
Because of these reasons, we take a sufficiently large value of $\zeta$.
Note that, for sufficiently large $\zeta$, $S$ is stabilized for any value of $T$.
We also assume $F(S)=0$ so that the potential value at the vacuum is zero.
Thus, $S$ is set to the minimum $S=0$, and
the resultant effective theory has two fields $\text{Re}T$ and $\text{Im}T$ with only one parameter $m$.

After stabilization of $S$, the Lagrangian density is given by
\begin{align}
\mc{L}=& - \frac{3}{\left(2 \text{Re}T+3\right)^{2}}\left( \partial_{\mu} \text{Re}T \partial^{\mu} \text{Re}T +\partial_{\mu} \text{Im}T \partial^{\mu} \text{Im}T \right) -\frac{3m^{2}}{\left(2 \text{Re}T+3\right)^{2}}\left( \text{Re}T^{2}+\text{Im}T^{2}\right).
\label{eq:ReTImT}
\end{align}
Canonical normalization of both fields at the same time is impossible
in this case.  We find it useful to define the semi-canonical basis
that does not have kinetic mixing and realizes canonical
normalization at the vacuum ($X=Y=0$):\footnote
{Alternatively, one may transform $\text{Im}T$ into canonically
  normalized form by $Z=\sqrt{\frac{2}{3}}e^{-\sqrt{2/3}X}\text{Im}T$.
  Then the potential for $Z$ is also simplified,
  $V=V^{\text{S}}(X)+\frac{m^{2}}{2}Z^{2}$, where $V^{\text{S}}(X)$ is
  the Starobinsky potential.  However, in this basis, $X$ is no more canonically
  normalized and there is a kinetic mixing between $X$ and
  $Z$.  } 
\begin{align}
X=\sqrt{\frac{3}{2}}\ln \left( 1+ \frac{2}{3}\text{Re}T  \right), \quad \quad 
Y=\sqrt{\frac{2}{3}}\text{Im}T.
\end{align}
Then, the Lagrangian density becomes
\begin{align}
\mc{L}=& - \frac{1}{2} \partial_{\mu} X \partial^{\mu} X -\frac{1}{2}e^{-2\sqrt{2/3}X}\partial_{\mu} Y \partial^{\mu} Y -\frac{3m^{2}}{4}\left(1-e^{-\sqrt{2/3}X}  \right)^{2}-\frac{m^{2}}{2}e^{-2\sqrt{2/3}X}Y^{2}.
\label{eq:LXY}
\end{align}
The third term is the Starobinsky potential.  Looking at the second
and fourth terms, one may naively guess that chaotic inflation~\cite{Linde:1983gd} is
possible neglecting the common factor $e^{-2\sqrt{2/3}X}$.  
However, as we shall see, 
this exponential factor strongly drives $X$ to the positive
direction in the large $Y$ region.

Now let us investigate if the fields $X$ and/or $Y$ play
the role of inflaton which are responsible for the present density
fluctuations of our universe.  For this purpose, we first study the
evolution of these fields.  The evolution equations for $X$ and $Y$
are given by
\begin{align}
&
\ddot{X} + 3 H \dot{X}
+
\sqrt{\frac{3}{2}} m^2 e^{-\sqrt{2/3}X} \left(1-e^{-\sqrt{2/3}X}\right)
-
\sqrt{\frac{2}{3}} e^{-2\sqrt{2/3}X}  \left(m^2 Y^2-(\dot{Y})^2 \right)
=0\,,
\\
&
\ddot{Y} +3H\dot{Y}
-2\sqrt{\frac{2}{3}}\dot{X}\dot{Y}
+ m^2 Y
=0\,,
\end{align}
where the ``dot'' denotes the derivative with respect to time $t$ and
$H\equiv \dot{a}/a$ (with $a$ being the scale factor) is the expansion
rate of the universe.  When the energy density of the universe is
dominated by that of $T$, we obtain
\begin{align}
  H = \sqrt{\frac{\rho_T}{3}}
  \end{align}
where $\rho_T$ is the total energy density:
\begin{align}
  \rho_T &= K_T+V_T\,,
\\
  K_T &= 
\frac{1}{2} \dot{X}^2
+
\frac{1}{2}e^{-2\sqrt{2/3}X} \dot{Y}^2\,,
\\
V_T &=
\frac{3m^2}{4} \left(1-e^{-\sqrt{2/3}X}\right)^2
+
\frac{m^2}{2} e^{-2\sqrt{2/3}X} Y^2\,.
\end{align}
By solving the above equations numerically, we follow the trajectories
of $X$ and $Y$ with various initial values.

\begin{figure}[t]
\begin{center}
 \includegraphics[width=10cm]{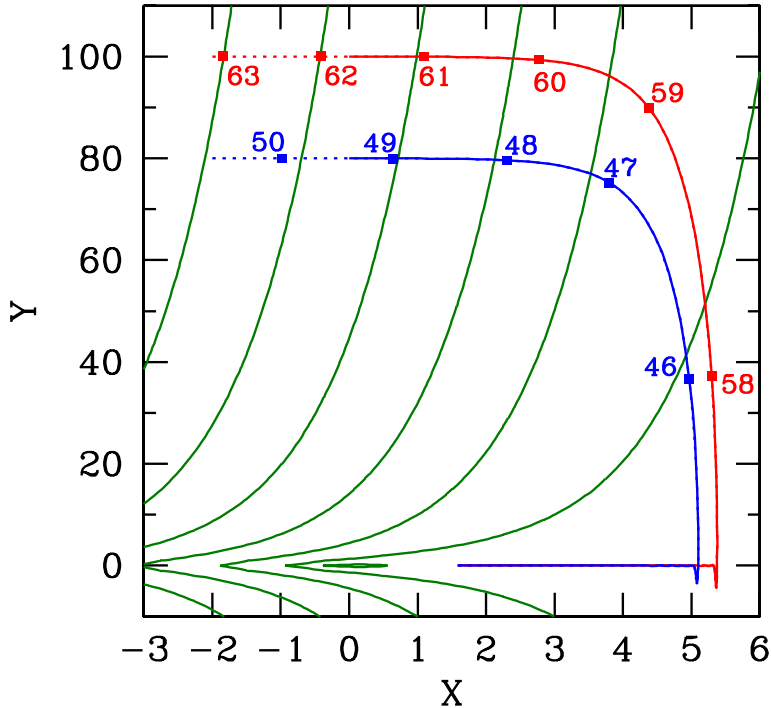}
 \end{center}
 \caption{Evolutions of the fields on the $(X, Y)$ plane.  The green
   closed contour at around the minimum ({\it i.e.}, $(X,Y)=(0,0)$) 
   corresponds $V_T(X,Y)/m^2=0.1$, while other green lines
   represent the contours of $V_T(X,Y)/m^2=1$, $10$,
   $10^2$, $10^3$, $10^4$ and $10^5$ from right to left.  The solid red,
   solid blue, dashed red, and dashed blue lines represent the
   evolutions of the fields with initial conditions $(X(t_{\rm
     init}),Y(t_{\rm init}))=(0,100)$, $(0,80)$, $(-2,100)$ and
   $(-2,80)$, respectively. Note that dashed lines overlap with the solid
   lines for $X>0$.  Points with numbers show the $e$-folding numbers
   for each trajectory.  The trajectories are terminated at the end of
   inflation ({\it i.e.}, $\epsilon_{H}=1$).
 }
\label{fig1:trajectory}
\end{figure}

In Fig.\ \ref{fig1:trajectory}, we show the contours of the potential
and the evolutions of the fields on $(X, Y)$ plane.  As representative
initial conditions, we choose $(X(t_{\rm init}),Y(t_{\rm
  init}))=(0,100)$, $(0,80)$, $(-2,100)$ and $(-2,80)$. (The initial
values of $\dot{X}$ and $\dot{Y}$ are taken to be zero.)  With such
initial conditions, we can see that $T$ starts to move to the $X$
direction first, then it settles to the real axis (i.e., $Y\simeq 0$).
After reaching to the real axis, the motion of $T$ is well
approximated by the single-field inflation with $X$; the situation is
almost the same as the non-supersymmetric original Starobinsky inflation.
As can be seen from the dashed lines, the trajectories are almost
unchanged even if $X$ starts from $X<0$.

On each contour, in particular for $Y\neq 0$, we show several points
which give rise to some specific values of the $e$-folding numbers
until the end of inflation.  Here, the $e$-folding number is defined
as
\begin{align}
  N_e (t)
  \equiv \int^{t_{\rm end}}_t dt' H(t'),
\end{align}
where $t_{\rm end}$ is the time at the end of inflation.  In our
analysis, we define it by $\epsilon_H(t_{\rm end})=1$, where the
slow-roll parameter $\epsilon_H$ is given by
\begin{align}
  \epsilon_H \equiv - \frac{\dot{H}}{H^2} = 
  1 - \frac{\ddot{a}}{a H^2}
  =\frac{3K_T}{\rho_T}\,.
\end{align}
We have used the Einstein equation in the last equality.  We can see
that the change of the $e$-folding value in the period of $Y\gg 1$ is
small.  Therefore, a large value of the $e$-folding number during
inflation, which is necessary to solve the horizon and flatness
problems, should be accumulated when $T$ is on the real axis.

\begin{figure}[t]
\begin{center}
 \includegraphics[width=10cm]{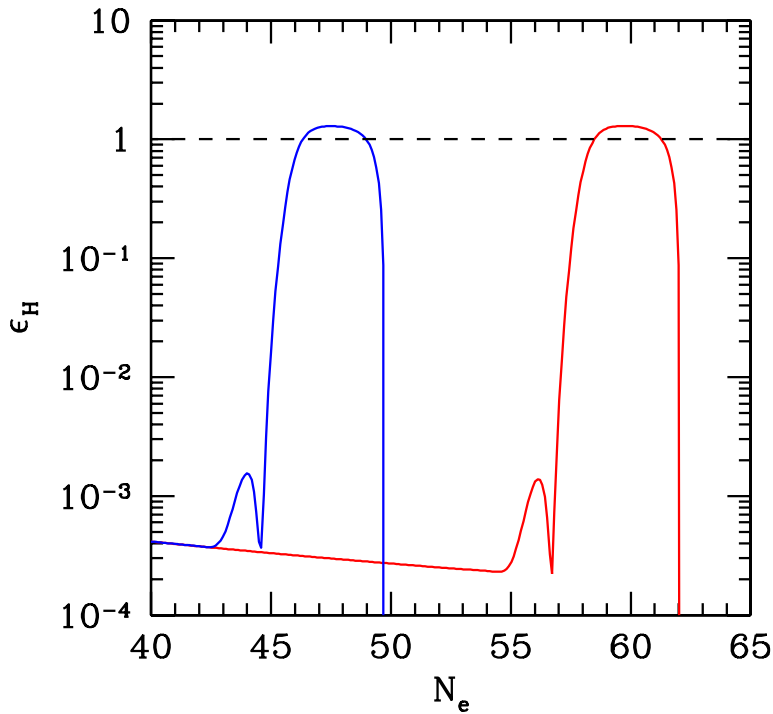}
 \end{center}
 \caption{The slow-roll parameter $\epsilon_H$ as a function of the
   $e$-folding number $N_e$, for the initial conditions $(X(t_{\rm
     init}),Y(t_{\rm init})) = (0,100)$ (red line) and $(X(t_{\rm
     init}),Y(t_{\rm init})) = (0,80)$ (blue line).  The dashed line,
   corresponding to $\epsilon_H=1$, is drawn to guide the eyes.  Note
   that $N_e$ decreases with time.  }
\label{fig2:epsH}
\end{figure}

For $\epsilon_H<1$ and $\epsilon_H>1$, the expansion of the universe is accelerating
and decelerating, respectively.  Thus, for inflation to happen,
$\epsilon_H<1$ is necessary.  To see when the expansion is accelerating, 
in Fig.\ \ref{fig2:epsH}, we
plot $\epsilon_H$ as a function of $N_e$, taking $(X(t_{\rm
  init}),Y(t_{\rm init})) = (0,100)$ and $(X(t_{\rm init}),Y(t_{\rm
  init})) = (0,80)$. 
  We can see that, just after the start of the
motion, $\epsilon_H$ significantly increases and soon becomes larger
than $1$. In this period, the expansion of the universe is decelerating and not
inflating.  The drop of $\epsilon_H$ at $N_e\simeq 57$ (45) in the red
(blue) line corresponds to the point at which $Y$ becomes most
negative and $\dot{Y}\simeq 0$ (cf. Fig.\ \ref{fig1:trajectory}).

Thus, the universe transits from the
decelerating epoch to the Starobinsky-type inflation.  We call the
period in between as ``transition period,'' and the period of the
Starobinsky-like expansion as ``Starobinsky-inflation period.''  The
important point is that the transition period is very short; during
the transition period, $N_e$ changes $\sim 3$ or so.  (For the case of
$(X(t_{\rm init}),Y(t_{\rm init})) = (0,100)$, for example, the
transition period corresponds to $55\lesssim N_e\lesssim 58$.)  This
is due to the fact that the motion of $Y$ becomes suppressed soon
after the condition $\epsilon_H<1$ is satisfied.  If we require
that the causal connection be realized for the scale much longer than
$k_*^{-1}$ (with $k_*$ being the wavenumber corresponding to the
present Hubble scale), the mode with the wavenumber $k_*$ should leave
the horizon in the Starobinsky-inflation period.  Then, the
tensor-to-scalar ratio becomes $O(10^{-3})$ and is too small to be
consistent with the value given in Eq.\ \eqref{r(BICEP2)}.  Thus, in
the light of the recent BICEP2 result, the Starobinsky inflation is
disfavored even if the field space is complexified in the framework of
old-minimal supergravity.

One of the possibilities to change this conclusion may be to consider
the case where the mode with $k_*$ exits the horizon in the transition
period.  However, such a solution looks unlikely.  Even though the
density fluctuations with the wavenumber $\sim k_*$ may be altered,
fluctuations with the wavenumber $k$ larger than $\sim 10 k_*$ have 
almost the same property as those in the case of Starobinsky
inflation.  Consequently, for the angular scale of
$\theta\lesssim\pi/l$ with $l\gtrsim O(10)$, the density perturbations
behave as those in the Starobinsky model.  The BICEP experiment is
sensitive to the $B$-mode signal with $l\sim 50-150$, while the
scalar-mode fluctuations for such an angular scale is well studied by
using CMB and other observables.  Thus, in the present model,
 it is difficult to enhance the tensor-mode fluctuations
without conflicting observations.

\section*{Note added}

While we are preparing the manuscript, the paper
\cite{Kallosh:2014qta} showed up on arXiv, which has some overlap with this letter.
See also Refs.~\cite{Ellis:2014rxa} and \cite{Ferrara:2014ima} for recent related works.

\section*{Acknowledgment}
This work was supported by JSPS KAKENHI Grant No.~22244021 (K.H.,
T.M.), No.~22540263 (T.M.), No.~23104008 (T.M.)  and also by World
Premier International Research Center Initiative (WPI Initiative),
MEXT, Japan. The work of T.T. was supported by the Program for Leading
Graduate Schools, MEXT, Japan.


\end{document}